\documentclass{aa} 
\usepackage{graphicx}
\usepackage{url}
\usepackage{natbib}
\usepackage{txfonts}
\usepackage[normalem]{ulem}
\usepackage{longtable}
\usepackage{color}
\usepackage{amssymb,xcolor}
\usepackage{amsmath}

\newcommand{\swift}  {Swift~J0243.6+6124}

\begin{document}

   \title{Unveiling the \textbf{origin} of the optical and UV emission \textbf{during the 2017 giant outburst of the Galactic ULX pulsar} Swift J0243.6+6124 (Corrigendum)}

   \subtitle{}
  \author{J. Alfonso-Garzón \inst{1} \and J. van den Eijnden \inst{2} \and N. P. M. Kuin \inst{3}  \and F. F\"{u}rst \inst{4} \and A. Rouco-Escorial \inst{5} \and J. Fabregat \inst{6} \and \\ P. Reig \inst{7,8} \and J. M. Mas-Hesse \inst{1} \and P. A. Jenke \inst{9} \and C. Malacaria \inst{10} \and C. Wilson-Hodge \inst{11}
}

\authorrunning{}
\titlerunning{\swift}
\offprints{}

   \institute{Centro de Astrobiolog\'{\i}a (CAB), INTA--CSIC, ESAC, Camino Bajo del Castillo s/n, 28692 Villanueva de la Ca\~nada, Spain \\ \email{julia@cab.inta-csic.es}
        \and Department of Physics, University of Warwick, Coventry CV4 7AL, UK
        \and Mullard Space Science Laboratory, Holmbury St Mary, Dorking, Surrey RH5 6NT, UK
        \and Quasar Science Resources SL for ESA, European Space Astronomy Centre (ESAC), 28692 Villanueva de la Can\~ada, Madrid, Spain
        \and European Space Agency (ESA), European Space Astronomy Centre (ESAC), Camino Bajo del Castillo s/n, 28692 Villanueva de la Ca\~nada, Spain
        \and Observatorio Astron\'omico, Universidad de Valencia, Catedr\'atico Jos\'e Beltr\'an 2, 46980 Paterna, Spain
        \and Institute of Astrophysics, Foundation for Research and Technology-Hellas, 71110 Heraklion, Greece
        \and Physics Department, University of Crete, 71003 Heraklion, Greece
        \and University of Alabama in Huntsville, Huntsville, AL 35805, USA  \and International Space Science Institute, Hallerstrasse 6, 3012 Bern, Switzerland
        \and ST12 Astrophysics Branch, NASA Marshall Space Flight Center, Huntsville, AL 35812, USA
        }

   \date{Received }

   \maketitle

\nolinenumbers

In the original paper, we showed the extraordinary optical/UV emission that \swift\, displayed during its 2017 giant outburst, and presented a physical model aimed to explain the effects of the X-ray irradiation on different elements in the system, including the heating of the companion star and the accretion disk. We estimated the X-ray luminosity for every day of the outburst and used it as an input to our model. In this erratum we want to point out a computing error in the calculations of one of the components of the model that led to an overestimate of the temperatures reached at the Be star surface due to the X-ray heating. As a consequence, the contribution of the irradiated Be star to the total optical/UV flux was overestimated by approximately an order of magnitude at the observed wavelengths. A reanalysis of the data has revealed that the emission by the Be decretion disk, which is also heated by the X-ray flux, must also be taken into account to properly reproduce the observations. The main conclusion of the original paper that the optical/UV emission is due to X-ray irradiation of cooler material in the system remains valid once the contribution by the Be decretion disk is considered.

\begin{figure}[h]
\begin{center}
\includegraphics[width=0.45\textwidth]{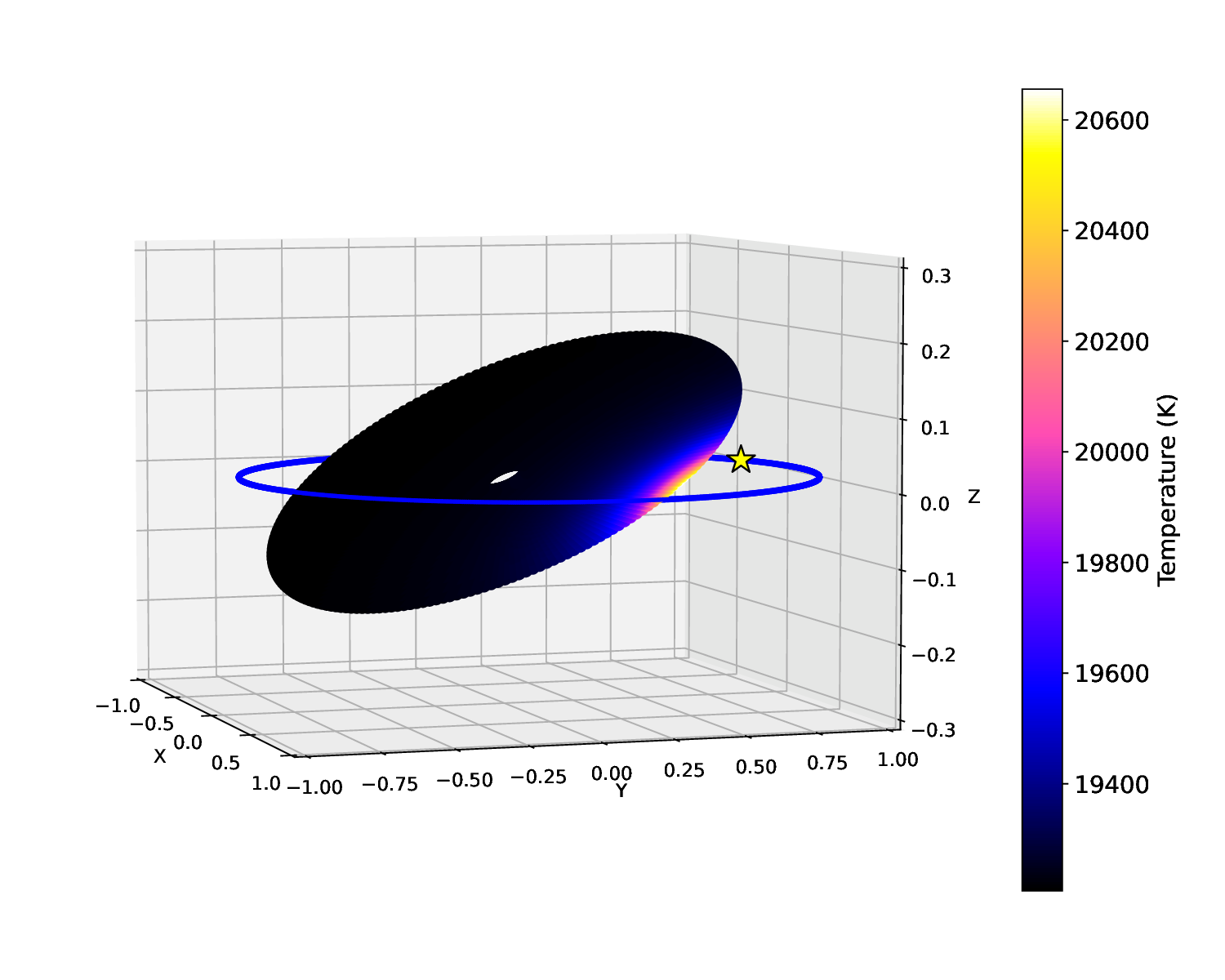}
\caption[]{Temperature distribution reached by X-ray heating of the Be disk at MJD\,58062 for a tilting angle $\phi$=15$^{\circ}$ (see text).}
\label{fig:Bedisk}
\end{center}
\end{figure}

One of the considerations we were making in the original paper was that the Be disk was co-planar to the orbit, and because of that, the contribution by the potential heating of the Be disk would be negligible and was not considered as a component of our model. However, recent polarimetric studies combining X-ray and optical observations indicate that the Be disk could be tilted with respect to the orbital plane during the 2023 outburst \citep{Poutanen2024}, and a similar behaviour for the 2017 outburst is expected. In this case, the irradiation of a misaligned Be disk should be considered. Preliminary calculations of such a scenario are providing promising results; in addition, a a detailed analysis considering a grid of parameters, such as the misalignment angle and the outer radius of the disk, is being performed and will be presented in an upcoming paper.

As an illustrative example, we show the results of the estimated temperatures reached at every surface element of an X-ray irradiated Be disk during the peak of the outburst, considering an initial set of parameters (Fig.\,\ref{fig:Bedisk}). As a first attempt, we have assumed for simplicity an isothermal disk with effective temperature T$_{\textnormal{eff,disk}}$=0.6$\times$T$_{\textnormal{eff,}\star}$ \citep{Carciofi2006}, with T$_{\textnormal{eff,}\star}$ being the effective temperature of the Be star. The radius of the disk has been set to R$_{\textnormal{disk}}$=10 R$_{\star}$, which corresponds to a Be disk that is filling its Roche lobe at the periastron passage (RL$_{\textnormal{Be}}$=0.8$\times a$), an albedo of 0.1, an angle of the Be disk misalignment with respect to the orbital plane $\phi$=15$^{\circ}$, and considering that the tilting axis is in the periastron - apastron direction. We have chosen this direction because for a circular Be disk, truncated at periastron passage, this is the simplest configuration to enable the material to be captured at the periastron passage and display X-ray activity. This value for the misalignment angle has been chosen as an intermediate value between those that, considering the orbital parameters of the system, can reproduce the differences between the polarization angle (PA) of the constant X-ray component ($\chi_{c}\sim$ 10$^{\circ}$, associated with the accretion disk and the orbit) and the intrinsic optical PA ($\chi_{o}$, related to the Be disk, which is in the range 20--50$^{\circ}$) given by \citet{Poutanen2024}.

\begin{figure}[h]
\begin{center}
\hspace{-0.5cm}\includegraphics[width=0.5\textwidth]{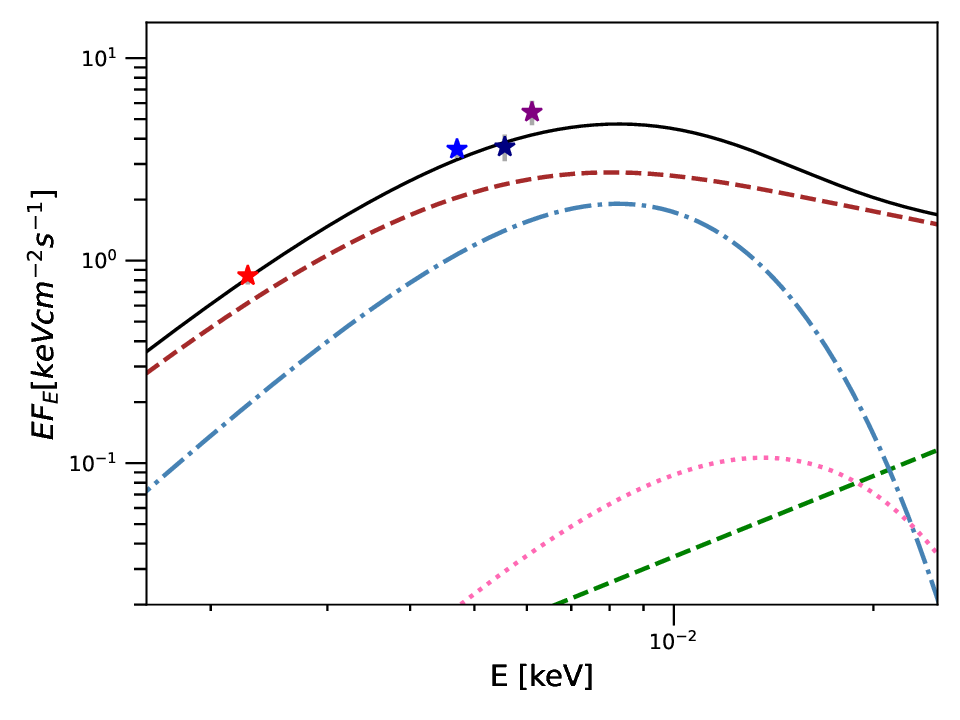}
\caption[]{Optical/UV SED at MJD\,58062 (close to the peak of the outburst). The colour code of the observations is the same as in Fig. 10 of the original paper. The dash-dotted blue line represents the contribution of the irradiated Be disk. The corrected contribution of X-ray heating of the companion star is plotted with a dotted pink line, the irradiated accretion disk is plotted as a dashed brown line, and the viscously heated accretion disk is plotted with a dash-dotted green line. The sum of the four components of the model is represented with a solid black line.}
\label{fig:SED_models}
\end{center}
\end{figure}

The de-reddened monochromatic fluxes in the $V$-Johnson (red star), UVOT-$uw1$ (blue star), UVOT-$um2$ (navy blue star), and UVOT-$uw2$ filters (purple star) from MJD\,58062 are shown in Fig.\,\ref{fig:SED_models}. The estimated contribution of the different components of the updated model, including the irradiated Be disk (dash-dotted blue line), the irradiated accretion disk (dashed brown line), the now correct irradiation of the Be star (dotted pink line), and the emission from the viscously heated accretion disk (dashed green line) are shown. The sum of the four components is represented with a solid black line.

A detailed analysis of the new model, including the effects of X-ray irradiation of a misaligned Be disk and considering a grid of different parameters, will be presented in a forthcoming paper.

\bibliographystyle{aa}
\bibliography{corr}

\end{document}